\documentclass{iaus}
\usepackage{graphicx}

\title[IAU Symp. 265~~Metallicity gradients in the Milky Way] 
{Metallicity gradients in the Milky Way}

\author[Walter J. Maciel \& Roberto D. D. Costa]   
{Walter J. Maciel  \and Roberto D. D. Costa}

\affiliation{University of S\~ao Paulo, Astronomy Department, \\ Rua do Mat\~ao 1226,
Cidade Universit\'aria, S\~ao Paulo SP, CEP 05509-0900, Brazil \\ email: {\tt maciel@astro.iag.usp.br}, 
{\tt roberto@astro.iag.usp.br}}

\pubyear{2009}
\volume{265}  
\pagerange{1--12}
\jname{Chemical abundances in the Universe: Connecting the first Stars to Planets}
\editors{K. Cunha, M. Spite, \& B. Barbuy, eds.}
\begin{document}

\maketitle

\begin{abstract}
Radial metallicity gradients are observed in the disks of the Milky Way and in several other
spiral galaxies. In the case of the Milky Way, many objects can be used to determine the 
gradients, such as HII regions, B stars, Cepheids, open clusters and planetary nebulae. Several 
elements can be studied, such as oxygen, sulphur, neon, and argon in photoionized nebulae, and 
iron and other elements in cepheids, open clusters and stars. As a consequence, the number 
of observational characteristics inferred from the study of abundance gradients is very large, 
so that in the past few years they have become one of the main observational constraints of 
chemical evolution models. In this paper, we present some recent observational evidences of 
abundance gradients based on several classes of objects.  We will focus on (i) the magnitude of 
the gradients, (ii) the space variations, and (iii) the evidences of a time variation of the
abundance gradients. Some comments on recent theoretical models are also given, in an effort 
to highlight their predictions concerning  abundance gradients and their variations.

\end{abstract}

\firstsection 
\section{Introduction}

Radial metallicity gradients are observed in the disks of many galaxies, including the Milky Way, 
galaxies of the Local Group and other objects. Current research topics include the determination 
of (i) the magnitude of the gradients, (ii) any space variations along the  disk and (iii) possible 
time variations during the evolution of the host galaxy. In this paper, we present some 
recent observational evidences of radial abundance gradients. Our main focus is the Milky Way, but 
it will be shown that the analysis of some objects in the Local Group, particularly M33, is  useful 
in order to study the main properties of the gradients in our own Galaxy. A brief discussion of some 
recent theoretical models is also given, in an effort to highlight their predictions concerning the 
radial abundance gradients and their variations. 
Some recent reviews and general papers on abundance gradients include: Freeman (\cite{freeman}), 
Maciel \& Costa (\cite{mc2009}), Rudolph et al. (\cite{rudolph}), and Stasi\'nska (\cite{stasinska04}). 
Theoretical models are discussed by a number of people, including Fu et al. (\cite{fu}),
Magrini et al. (\cite{magrini09a}), Cescutti et al. (\cite{cescutti}), Moll\'a and D\'\i az 
(\cite{molla}), Chiappini et al. (\cite{chiappini03}, \cite{chiappini01}, \cite{chiappini97}),  
Hou et al. (\cite{hou}), and Gensler (these proceedings). Recent discussions on azimuthal and 
vertical gradients, not treated here, can be found in Davies et al. (\cite{davies}) and Ivezi\'c 
et al. (\cite{ivezic}).

\section{Abundance gradients in the Milky Way}

\subsection{Cepheids} 

Cepheid variables are possibly the most accurate indicators of abundance gradients in the Milky Way. 
Since the work of Andrievsky and collaborators  (cf. Andrievsky et al. \cite{sergei1}, \cite{sergei2}, 
\cite{sergei3}, \cite{sergei4}, Luck et al. \cite{luck}), several papers have dealt with these objects 
in order to study not only the magnitudes of the present-day gradients, but also the detailed behaviour 
of the abundances along the galactic disk. The main reason is that cepheids are bright enough to be 
observed at large distances, so that accurate distances and spectroscopic abundances of several elements 
can be obtained, which in principle allow the determination of radial variations of the gradients better 
than any other indicator. The ages of these objects are generally under 200 Myr (cf. Maciel et al. 
\cite{mlc2005}), so that the measured gradients can be safely considered as present-day gradients.
Recent work on Cepheids include Lemasle et al. (\cite{lemasle08}), Pedicelli et al. (\cite{pedicelli}), 
and Romaniello et al. (\cite{romaniello}). Lemasle et al. (\cite{lemasle08}) obtained high-resolution 
spectroscopic iron abundances for galactic cepheids for which accurate distances were determined based 
on  near-infrared photometry.  The abundances were determined within 0.12 dex, while the distances, 
based on a near-infrared period-luminosity relation, are expected to be accurate within 0.5 kpc in average. 
In order to improve  the sampling process, additional objects were included. The  average gradient obtained 
in the range 5--17 kpc is $d[{\rm Fe/H}]/dR = -0.052 \pm 0.003$ dex/kpc. A better solution proposed by 
the authors includes a change of slope, in the sense that the gradient is steeper in the inner galaxy, 
with a flattened gradient of $-0.012 \pm 0.014$ dex/kpc in the outer galaxy. In this region the abundances 
show an increased spread, as compared with the inner galaxy.  The change of slope occurs at about $R \simeq 
10$ kpc, farther away than the solar radius, located at  $R = 8.5$~kpc. These results have been largely 
confirmed by the recent work of Pedicelli et al. (\cite{pedicelli}), in which again a large sample from 
different sources was considered with a new photometric metallicity calibration. These results suggest a 
gradient of $-0.051 \pm 0.004$ dex/kpc for the whole sample, or $-0.130 \pm 0.015$ dex/kpc for the inner 
sample ($R < 8$ kpc) and $-0.042 \pm 0.004$ dex/kpc for the outer Galaxy.  

\subsection{HII regions}

Concerning HII regions, several determinations have been presented in the last few years, which 
include Deharveng et al. (\cite{deharveng}),  Esteban et al. (\cite{esteban}), Rudolph et al. 
(\cite{rudolph}) and Quireza et al. (\cite{quireza}). Deharveng et al. (\cite{deharveng}) obtained 
an oxygen gradient of $-0.04$ dex/kpc in the range 5--15 kpc, with no indication of flattening further 
out in the disk. Pilyugin et al. (\cite{pilyugin}) compiled spectra of 13 HII regions in the range 
7--14 kpc with available [OIII]$\lambda$4363 measurements, and recomputed the oxygen abundances from 
the data by Shaver et al. (\cite{shaver}), obtaining a gradient of $-0.051$ dex/kpc. Esteban et al. 
(\cite{esteban}) obtained echelle spectrophotometry of 8 HII regions 
in the range 6--10 kpc, and derived carbon and oxygen abundances from recombination lines. The oxygen 
gradient obtained is $-0.044 \pm 0.010$  dex/kpc. More recently, Rudolph et al. (\cite{rudolph}) used 
both infrared and optical data to study abundance gradients in a sample of 117 HII regions. The data 
include both new results and a reanalysis of previous material. The best fit to optical data corresponds 
to a gradient of $-0.060$ dex/kpc. For the infrared data a gradient of  $-0.041$ dex/kpc was obtained.  
Quireza et al. (\cite{quireza}), determined the electron temperature gradient from high-precision radio 
recombination line and continuum measurements for over a hundred HII regions, calibrated in  terms of 
the oxygen abundance gradient. The derived slope obtained using a relation between the electron temperature 
and the oxygen abundances by Shaver et al. (\cite{shaver}) is  $-0.043 \pm 0.007$ dex/kpc, in good 
agreement with the previous work by Deharveng et al. (\cite{deharveng}) and Esteban et al. (\cite{esteban}).

From these results it is difficult to establish whether or not the gradients flatten out at large 
galactocentric distances.  However, studies of HII regions in spiral galaxies are consistent with essentially 
constant abundances in the outer parts of these objects, as for example in the recent work of Bresolin 
et al. (\cite{bresolin09a}) on the extended disk of M83, in which a flat oxygen gradient was obtained beyond 
the R25 isophotal radius, regardless of the abundance indicator used. 

\subsection{Stars and Open Clusters} 

Apart from Cepheids, other field stars  can be used in order to analyze the abundance gradients in the 
Milky Way.  For open cluster stars, both the metallicity and the distances are well determined, and they 
have a reasonably large age span, so that they can be used to investigate both the spatial and temporal 
changes in the gradients. However, the age determinations  may depend on the calibration used. Work up to 
2007 is summarized by Cescutti et al. (\cite{cescutti}), where gradients of O, Mg, Si, S, and Ca are 
discussed. Several objects have been considered, comprising Cepheids,  O, B stars, red giants, and two 
samples of open clusters.  The main feature  which is common to all data is a change of slope at large 
galactocentric radii, roughly $R > 10$ kpc, which characterizes the flattening of the gradients in the 
outer Galaxy. Theoretical models are also discussed, which assume an inside-out formation for the 
galactic disk with a time scale of 7 Gyr for the thin disk in the solar vicinity and a shorter timescale 
of 0.8 Gyr for the galactic halo. The inside-out scenario is apparently a necessary condition to explain 
the present-day gradients and the variation of the star formation rate with the galactocentric radius. 
In fact, recent models by Colavitti et al. (\cite{colavitti09}) conclude that all disk constraints  
cannot be simultaneously satisfied unless an inside-out formation of the galactic disk is assumed.

More recent work on open clusters (Sestito et al. \cite{sestito}, Magrini et al. \cite{magrini09a}) 
generally confirm these findings, using larger samples. A steep negative gradient is found for the inner 
Galaxy up to about 10--11 kpc, with a flat distribution in the outer Galaxy. A comparison is made of these 
results with the earlier work by Friel et al. (\cite{friel}) based on low-resolution spectroscopy. Although 
the error bars are larger, there is not much difference relative to the high-resolution data for the inner 
Galaxy. The slopes are $-$0.17 dex/kpc and $-$0.09 dex/kpc for these samples, considering only the inner region. 
It is doubtful whether the difference is meaningful at this stage. In the outer Galaxy, an essentially flat 
gradient is observed for the new sample.

In view of the age span of the open clusters, they are ideally suited to study any time variations 
of the gradients. Magrini et al. (\cite{magrini09a}) considered a sample of open clusters with 
high-resolution data, in the range 7--22 kpc, with estimated ages in the range 30 Myr to 11 Gyr.  
The authors present plots of the gradients according to the age interval of the clusters for Fe, Cr, Ni, 
Si, Ca, and Ti, with a generally similar behaviour. Also, the ratios of these elements to Fe are essentially 
constant, suggesting that the derived gradients are similar within the uncertainties. Again, a steep gradient 
was found for the inner Galaxy ($R < 12$~kpc), with a flattening outwards. The flattening can be observed 
in all age brackets, but is especially clear in the sample with 4--11 Gyr.  Considering the gradients 
obtained by adopting three different galactocenctric ranges leads to the  conclusion that the gradients are 
approximately constant or slightly flattening with time. However, from the slopes of the inner Galaxy, 
this conclusion is probably a conservative one, as the indication of a flattening of the gradients seems clear. 

Recent large surveys  of galactic stars are also being used to derive abundance gradients, mainly based
on the [Fe/H] ratio. Some examples can be seen in the presentations by B. Nordstr\"om on the Geneva-Copenhagen
survey and by C. Boesche on the RAVE project (cf. the conference site of {\it The Milky Way and the Local
Group: Now and in the GAIA Era}, Heidelberg, 2009).

\subsection{Planetary Nebulae}

   \begin{figure}
   \centering
   \includegraphics[angle=0,width=12cm]{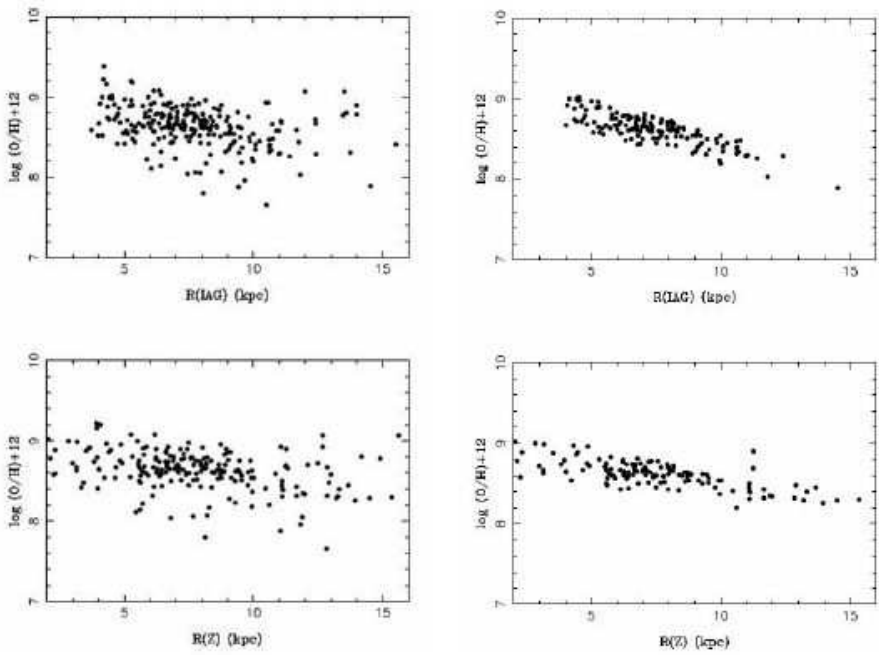}
       \caption{The O/H radial gradient from PN. Top: Distances from the IAG/USP
       group, Bottom: Distances from Stanghellini et al. (\cite{ssv}). Left: CSPN 
       with ages in the range 2--10 Gyr; Right: CSPN with ages in the range 4--6 Gyr.}
       \label{fig1}
   \end{figure}

The analysis of abundance gradients from planetary nebulae (PN) is hampered by some aspects related to 
these objects:  first, the distances to the galactic nebulae are often uncertain, and statistical scales 
have to be used in order to have a sizable sample; second, the progenitors of the planetary nebulae have 
a wide age span, as in the case of the open clusters, ranging from about 1 Gyr to about 8 Gyr (cf. Maciel 
et al. \cite{mci2009}), so that any time variation of the gradients would have to be taken into account. 
On the other hand, abundances of elements such as O, Ne, S, and Ar can be obtained within about 0.2 dex in 
average, which is probably lower than the abundance spread at a given galactocentric radius. The results 
presented in the last couple of years have been rather contradictory.  Pottasch and collaborators (cf. 
Pottasch and Bernard-Salas \cite{pottasch}) have presented accurate abundance data based both on optical 
and infrared measurements from ISO, which do not need the consideration of the usually uncertain ionization 
correction factors (ICF), since more ionized species can be observed. As a result, the derived abundances 
are expected to be more accurate than in the case of the traditional plasma diagnostic method. The results 
suggest the presence of a strong negative gradient similar to the ones observed in HII regions and early 
type stars. For the O/H ratio a slope of $-0.085$ dex/kpc was found. For most of the elements considered, 
which are O, Ne, S, and Ar, the predicted abundances at the solar radius obtained by taking into account 
the observed gradients  match exactly the solar abundances.  In a more recent work (Gutenkunst et al. 
\cite{gutenkunst}), a larger sample was considered, including bulge nebulae. It can be concluded that the 
gradient flattens out near the galactic bulge, a result also obtained by Cavichia et al. (\cite{cavichia}). 
In contrast with these results, Stanghellini et al.  (\cite{stanghellini06}) studied a sample of galactic 
PN with abundances derived from the traditional optical plasma diagnostic method, obtaining a flat gradient 
for oxygen and neon, based on a simple linear fit. However, from what we have seen in the previous sections, 
it seems clear that the gradients flatten out in the outer Galaxy, so that using a single fit for 
the whole disk may be misleading, especially if the inner Galaxy, where the gradient is steeper, is 
undersampled as is the case here. Moreover, the data in this sample located in the region around the solar 
circle clearly show some evidence of a steeper gradient, so that a flat gradient for the whole disk is probably 
incorrect. A more detailed analysis was made by Perinotto and Morbidelli (\cite{perinotto}), who considered a 
larger and more complete sample of PN for which the abundances were recalculated in a homogeneous way. The 
results also suggest relatively flat gradients ($<$ $-$0.04 dex/kpc) for oxygen, but a careful analysis of 
their data shows that the uncertainties in the distances, coupled with a possible time variation of the 
gradients, may wash out the gradients, so that a careful selection of the objects must be made. This can be 
seen by considering  the oxygen gradients for PN of sets A and B, defined as follows: set A includes 131 
objects whose abundances are considered by the authors as the most reliable, and set B is a control sample, 
containing all PN abundances published between 2000 and 2005, with about 200 objects. In order to avoid 
any bias due to the adopted distances, Perinotto and Morbidelli (\cite{perinotto}) considered four different 
statistical scales. Considering for example the results corresponding to the distances by Cahn et al. 
(\cite{cahn}), a gradient is apparent from set A, while set B presents a flat distribution, suggesting that 
the uncertainties in the abundances contribute to erase any existing gradients. Some indication of the 
existence of a time variation of the gradients, in the sense that the present-day gradient is flatter than 
in the past can also be observed from the results by Perinotto and Morbidelli (\cite{perinotto}). They have 
analyzed separately the PN according to the Peimbert types, in which Type I are expected to be younger objects, 
while Type III are older nebulae, generally located at higher distances from the galactic plane and with a 
larger peculiar velocity. Type II are intermediate age objects in this scheme. According to Perinotto and 
Morbidelli (\cite{perinotto}), Type I objects do not show any gradients in both sets, while Type II and III 
show measurable, albeit low, gradients. Also, for the distance scales with a meaningful sample, the gradients 
of Type III objects are larger than for Type II PN.

   \begin{figure}
   \centering
   \includegraphics[angle=0,width=8cm]{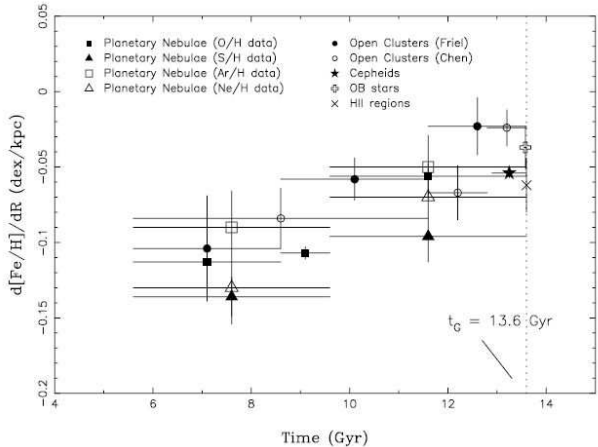}
       \caption{Time variation of the radial abundance gradient (Maciel \& Costa 
       \cite{mc2009}).}
       \label{fig2}
   \end{figure}

A different approach has been taken by the IAG/USP group (Maciel et al. \cite{mcu2003}, \cite{mlc2005}, 
\cite{mlc2006}, \cite{mci2009}, Maciel \& Costa \cite{mc2009}). Here an effort has been made to divide 
the PN sample into age groups, as was done for the open clusters by Magrini et al. (\cite{magrini09a}), 
so that any time variation of the gradients could be appropriately taken into account. Several methods
have been developed to obtain the age distribution of the PN central stars. Fig.~1 shows the 
O/H gradient for disk objects with ages of 2--10 Gyr (left) and 4--6 Gyr (right), using the distance
scale adopted by our IAG Basic Sample (top), or the distance scale by Stanghellini et al. (\cite{ssv})
(bottom). The effect of restricting the age interval is clear in both figures. Taking into account the age 
distribution of the PN progenitor stars (Maciel et al. \cite{mci2009}), it is possible to separate the PN 
sample according to their ages (young, intermediate, old), so that an estimate of the time variation of 
the gradients can be obtained. This is shown in Fig.~2 (Maciel \& Costa \cite{mc2009}), where other objects 
are also considered, namely open clusters, cepheids, OB stars and HII regions. As a conclusion, data from 
planetary nebulae support the flattening of the gradients near the bulge-disk interface and at large 
galactocentric distances. However, anticentre nebulae are difficult to observe, and the  problem of the 
distances is still a complicating factor. A considerable improvement is expected with  the advent of GAIA. 
As for the time variation of the gradients, a conservative conclusion at this stage is that either they 
have not changed very much in the last 6 Gyr approximately,  or they may have flattened out by a small 
amount. These conclusions are supported by some recent theoretical models  by Fu et al. (\cite{fu}), 
which take into account infall, star formation based on the Kennicutt law and a delayed disk formation.

\section{The Electron Temperature Gradient}

An important confirmation of the abundance gradients in photoionized nebulae comes from the expected 
electron temperature gradient, since the heavy elements for which radial gradients are observed are 
the main coolants in these objects. Therefore, a smilar, albeit inverted gradient is expected for HII 
regions and PN. This is in fact observed, as can be seen even in the earlier papers on this subject. 
Also, well defined electron temperature gradients have been measured in spiral galaxies such as NGC 300 
and  M101 (Bresolin et al. \cite{bresolin09b}) and M33 (Magrini et al. \cite{magrini07a}). More recently, 
Quireza et al. (\cite{quireza}) have determined accurate electron temperature gradients from radio 
recombination lines. The average HII region gradient is $287 \pm 46$ K/kpc. A somewhat higher gradient 
of $373$ K/kpc was obtained earlier by Deharveng et al. (\cite{deharveng}). Similar conclusions have 
been obtained by Maciel et al. (\cite{mqc2007}) for planetary nebulae, for which a steeper electron 
temperature gradient was derived amounting to about $670$ K/kpc. Fig.~3 shows both the electron temperature 
gradient of PN and the corresponding correlation with the oxygen abundances. The difference between the 
electron temperature gradients of HII regions (flatter) and planetary nebulae (steeper) is a strong 
indication that the gradients have flattened out since the PN progenitor stars have formed.

   \begin{figure}
   \centering
   \includegraphics[angle=-90,width=13cm]{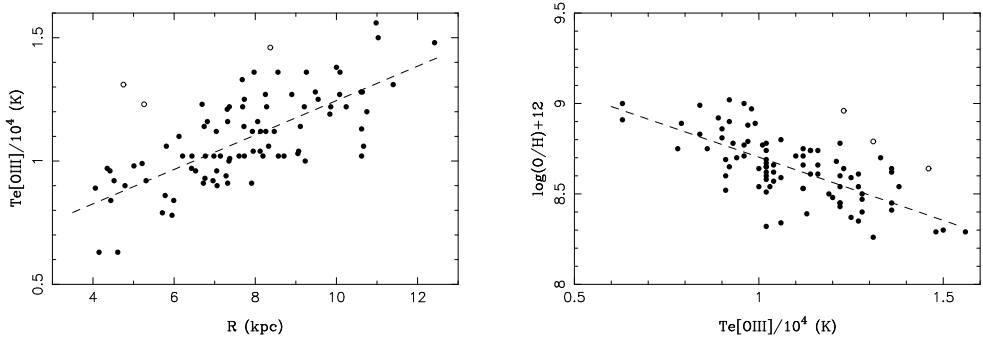}
       \caption{Left: The [OIII] electron temperature gradient from planetary nebulae.
       Right: The correlation between the electron temperatures and oxygen abundances 
       (Maciel et al. \cite{mqc2007}).}
       \label{fig3}
   \end{figure}

\section{ M33: A very interesting Case}

The galaxy M33 is a very interesting case, where abundance gradients have been recently measured both 
from HII regions and PN, apart from other objects. Rubin et al. (2008) have obtained 
Ne and S abundances in a sample of HII regions in M33 using Spitzer data, covering a wide range of 
galactocentric distances. Average gradients of $-0.058 \pm 0.014$ dex/kpc and  $-0.052 \pm 0.021$ dex/kpc 
are obtained for Ne/H and S/H, respectively. Magrini et al. (\cite{magrini07a}, \cite{magrini07b}) 
analyzed abundances of O, N, and S in a sample of HII regions and derived both electron temperature and 
abundance gradients for this galaxy within a radius of about 7 kpc from the galactic nucleus. The 
electron temperature gradient was also measured, amounting to about 570 K/kpc, corresponding to an 
O/H gradient of $-0.054$ dex/kpc. By considering additional objects from the recent literature, it is 
concluded that the oxygen data cannot be fitted with a single slope, so that an inner slope of $-0.19$ 
dex/kpc  and an outer slope of $-0.04$ dex/kpc are suggested. Theoretical models have been developed in 
which a continuous infall of gas on the disk is assumed. The models are calculated at different epochs, 
varying from an age of 2 Gyr to the present day, at 13.6 Gyr. According to this model, the gradients show 
some mild flattening with time. More recently, Magrini et al. (\cite{magrini09b}) considered a larger PN 
sample and obtained a relatively weak O/H gradient of $-0.03$ dex/kpc to $-0.04$ dex/kpc. Similar values 
were measured for Ne/H ($-0.03$ to $-0.05$ dex/kpc) and S/H ($-0.03$ to $-0.04$ dex/kpc). The recalculated 
O/H gradient for HII regions is similar to these values. Therefore, at face value the PN gradients are 
marginally steeper than the corresponding HII region gradients. However, a single slope was obtained for 
the whole sample, suggesting that the derived gradients are probably lower limits for the inner galaxy, 
since the gradients tend to flatten out at larger galactocentric distances. Moreover, PN may have progenitors 
with different ages, so that mixing these objects would contribute to flatten the measured slopes. This 
is reinforced by the fact that the inner sample, which is associated with the highest metallicities, is 
undesampled. A hint on this point can be obtained considering the results by Cioni (\cite{cioni}) on M33. 
She has considered a sample of well measured AGB stars, showing that the [Fe/H] gradient is clearly steepeer 
in the inner parts of the galaxy, flattening out at the outer parts. Since PN and AGB stars are objects of 
similar ages, their gradients are expected to be similar, which reinforces the conclusion that the 
flatter gradients found by Magrini et al. (\cite{magrini09b}) are lower limits.

\section{Conclusions}

From all objects considered, some tentative conclusions may be drawn: (1) Average abundance gradients 
are generally between $-0.03$ dex/kpc and $-0.10$ dex/kpc, but a single value for the whole disk may be 
misleading. (2) Most evidences point to a flattening  out of the gradients  at large galactocentric 
distances. (3) There are some clear evidences of a flattening  of the gradients  near the galactic bulge. 
(4) The change of slope in the outer Galaxy occurs in the region around $R \simeq 10$ kpc. (5) Any further 
change of the slope needs better data than presently available.  Cepheids may be an exception. (6) There 
are no evidences of a steepening of the gradients at large galactocentric distances, as suggested  by 
some theoretical models. (7) Either the gradients do not change appreciably during galactic evolution, or 
they flatten out at a moderate rate. (8) There are no clear evidences of a steepening of the gradients 
with time, as suggested by some theoretical models.

\bigskip\noindent
{\it Acknowledgements. This work was partially supported by FAPESP and CNPq.}

\end{document}